\documentclass[aps,prl,showpacs,twocolumn,floatfix,superscriptaddress]{revtex4} 
\usepackage{graphicx}

\begin{document}

\title{Large-scale collective properties of self-propelled rods} 
  
\author{Francesco Ginelli} 
\affiliation{Service de Physique de l'\'Etat Condens\'e, CEA--Saclay,~91191~Gif-sur-Yvette,~France} 
\affiliation{Institut des Syst\`emes Complexes de Paris \^Ile-de-France, 57-59 rue Lhomond, 75005 Paris, France}%
 
\author{Fernando Peruani} 
\affiliation{Service de Physique de l'\'Etat Condens\'e, CEA--Saclay,~91191~Gif-sur-Yvette,~France} 
 
\author{Markus B\"ar} 
\affiliation{Physikalisch-Technische Bundesanstalt, Abbestrasse 2-12, 
10587 Berlin, Germany} 
 
\author{Hugues Chat\'e} 
\affiliation{Service de Physique de l'\'Etat Condens\'e, CEA--Saclay,~91191~Gif-sur-Yvette,~France}

\date{\today} 
 
\begin{abstract}
We study, in two space dimensions,
the large-scale properties of collections of constant-speed polar 
point particles interacting locally by nematic alignment in the presence of noise.
This minimal approach to self-propelled rods allows one to deal with large 
numbers of particles, 
revealing a phenomenology previously unseen in more complicated models,
and moreover distinctively different from both that of the purely polar case (e.g. the 
Vicsek model) and of active nematics. 
\end{abstract}

\maketitle 


 
Collective motion is an ubiquitous phenomenon observable at all scales,   
in natural systems \cite{birdfish} as well as human societies \cite{humancrowds}.  
The mechanisms at its origin can be remarkably varied. For instance, they may involve 
the hydrodynamic interactions mediated by the fluid in which bacteria 
swim~\cite{sokolov2007}, the long-range
chemical signaling driving the formation and organization of aggregation centers 
of Dictyostelium discoideum amoeba cells~\cite{benjacob2000}, 
or the local cannibalistic interactions between marching locusts~\cite{romanczuk2009}.
In spite of this diversity, one may search for possible universal features of
collective motion, a context in which the study of ``minimal'' models is a crucial step.
Recently, the investigation of the simplest cases, where 
the problem is reduced to the competition between  
a local aligning interaction and some noise, 
has revealed a wealth of unexpected collective properties.
For example, constant speed, self-propelled, {\it polar} point particles 
with {\it ferromagnetic} interactions subjected to noise 
(as in the Vicsek model~\cite{Vicsek1995}) 
can form a collectively moving fluctuating phase with
long-range polar order even in two spatial dimensions~\cite{Toner}, 
with striking properties such as spontaneous segregation 
into ordered solitary bands moving in a sparse, disordered sea, or
anomalous (``giant'') density fluctuations \cite{Chate2008}. 
In contrast, active {\it apolar} particles with {\it nematic} interactions 
only exhibit quasi-long-range nematic order in two dimensions
with segregation taking the form of a single, strongly-fluctuating,
dense structure with longitudinal order
and even stronger density fluctuations than in the polar-ferromagnetic case
\cite{Chate2006,ramaswamy2006,Ramaswamy2003}. 
 
\begin{figure}[t] 
\centering 
\resizebox{\columnwidth}{!}{\rotatebox{0}{\includegraphics{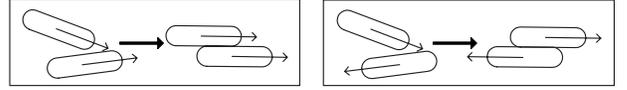}}} 
\caption{Nematic alignment of polar particles illustrated by inelastic collisions of
rods. Particles incoming at a small angle (left) align ``polarly'', but those
colliding almost head-on slide past each other, maintaining their nematic alignment
(right).} 
\label{fig:1} 
\end{figure} 
 
Noting that these differences reflect those in the local symmetry of  
particles and their interactions, a third situation can be defined, intermediate  
between the polar ferromagnetic model and the apolar nematic one, that of
self-propelled {\it polar particles aligning nematically} \cite{NOTE1}.
Such a mechanism is typically induced by volume exclusion interactions, 
when elongated particles colliding almost head-on slide past each other, 
as illustrated schematically in Fig.~\ref{fig:1}.
Thus, self-propelled polar point particles with apolar interactions can be conceived 
as a minimal model for self-propelled rods interacting by inelastic 
collisions~\cite{peruani2006, kudrolli2008,wensink}.
Other relevant situations can be found in a biological context, such as 
gliding myxobacteria moving on a substrate~\cite{myxo}, 
or microtubules driven by molecular motors grafted on a surface~\cite{filaments}.
 
\begin{figure*}[!t] 
\begin{center} 
\includegraphics[clip,width=18cm]{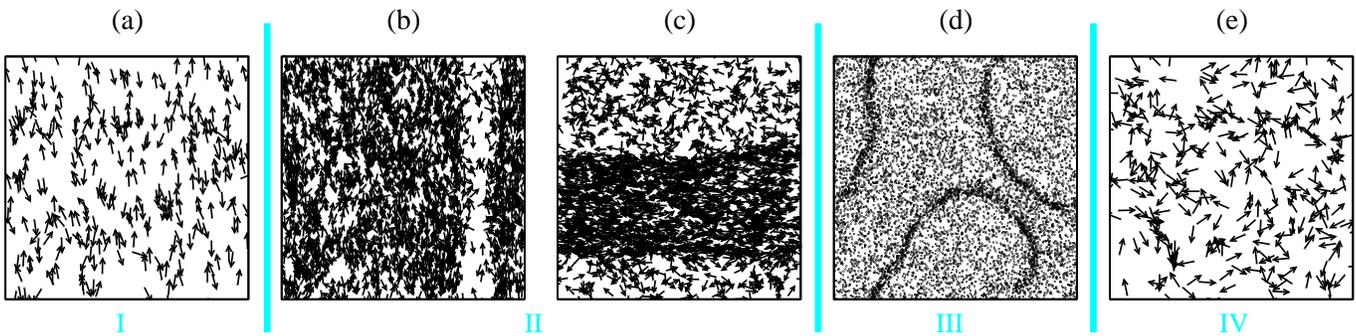} 
\caption{(color online) (a-c)
Typical steady-state snapshots at different noise values 
(linear size $L=2048$). 
(a) $\eta=0.08$, (b) $\eta=0.10$, (c) $\eta=0.13$, (d) $\eta=0.168$, (e) $\eta=0.20$. 
Arrows indicate the polar orientation of particles (except in (d)); only a fraction of  
the particles are shown for clarity reasons. For a movie corresponding to (d) see \cite{EPAPS}.}
\label{fig:2} 
\end{center} 
\end{figure*} 

In this Letter, we study collections of constant-speed polar 
point particles interacting locally by nematic alignment in the presence of noise. 
The simplicity of this model allows us to deal with large numbers of particles, 
revealing a phenomenology previously unseen in more complicated models sharing 
the same symmetries~\cite{peruani2006, kudrolli2008, wensink}. 
Our study, restricted to two space dimensions, shows in particular 
collective properties distinctively different from both those of 
polar-ferromagnetic case and of active nematics: only nematic order arises 
in spite of the polar nature of the particles, but it seems genuinely long-ranged.
Spontaneous density segregation is also observed but it is of a different type
and it splits both the  
(nematically) ordered and the disordered phase in two. 
In the following, we characterize these four phases and discuss
the three transitions separating them.
 
Our model consists of $N$ point particles moving off-lattice at constant speed $v_0$. 
In two dimensions, particle $j$ is defined by its  
(complex) position $\mathbf{r}_j^t$ and  
orientation $\theta_j^t$, updated at discrete time steps according to 
\begin{eqnarray} 
\label{motion_angle} 
\theta_j^{t+1}&=&  
\arg \left[\sum_{k\sim j}{\rm sign}\left[\cos(\theta_k^t -\theta_j^t)\right]  
e^{i\theta_k^t}\right] + \eta\xi_{j}^{t} \\  
\mathbf{r}_j^{t+1}&=& \mathbf{r}_j^{t}+v_0\, e^{i\theta_k^{t+1}} \, , 
\label{motion_pos} 
\end{eqnarray} 
where the sum is taken over all particles $k$ within unit distance of $j$  
(including $j$ itself), and 
 $\xi$ is a white noise uniformly distributed in  
$\left[-\frac{\pi}{2},\frac{\pi}{2}\right]$~\cite{NOTE2}. 
(A continuous-time version of this model can be found 
in \cite{peruani2008}.)
The system has two main control  
parameters: the noise amplitude $\eta$, and the particle density 
$\rho=N/A$, where $A$ is the domain area.  
We consider periodic boundary conditions. 
Polar and nematic order can be characterized by means of the two time-dependent 
global scalar order parameters $P(t)=|\langle\exp (i\theta^t_j)\rangle_j|$  
(polar) and $S(t)=|\langle\exp (i 2\theta^t_j)\rangle_j|$ (nematic), 
as well as their asymptotic time averages $P=\langle P(t)\rangle_t$ 
and $S=\langle S(t)\rangle_t$. 

\begin{figure}[b] 
\begin{center} 
\includegraphics[clip,width=8.6cm]{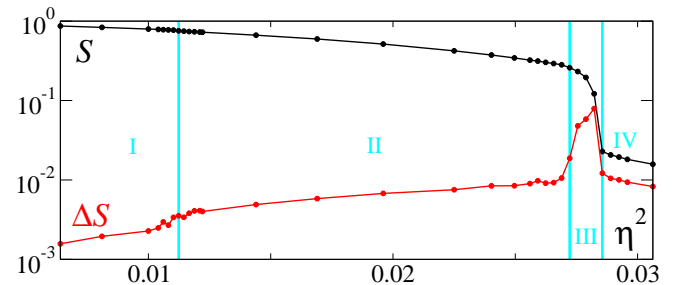} 
\caption{(color online) Nematic order parameter $S$ (in black) 
and its rms fluctuations $\Delta S$ (in red) as function of the 
squared noise amplitude $\eta^2$ for a square domain of linear size $L=2048$. 
Here, and throughout the paper, time-averages are over at least $10^6$ timesteps.} 
\label{fig:3} 
\end{center} 
\end{figure}

In this work, we focus on the behavior of 
the system for $\rho=\frac{1}{8}$ and $v_{\rm 0}=1/2$, varying $\eta$. 
We start with a brief survey of the stationary states 
observed in a square domain of linear size $L=2048$ (Figs.~\ref{fig:2}-\ref{fig:3}).
Despite the polar nature of the particles, only {\it nematic} orientational 
order arises at low noise strengths,
while $P$ always remains near zero (not shown).
This is in agreement with the findings of \cite{baskaran2008}. 
Both the ordered and the disordered regimes are 
subdivided in two phases, one that is spatially homogeneous (Figs.~\ref{fig:2}(a,e)), 
and one where spontaneous density segregation occurs,
leading to high-density ordered bands along which the particles move back and forth 
(Figs.~\ref{fig:2}(b-d)).
A total of four phases is thus observed, labeled I to IV by  
increasing noise strength hereafter. Phases I and II are nematically ordered, 
phases III and IV are disordered.
Below, we study these four phases more quantitatively.
 
\begin{figure}[t] 
\begin{center} 
\includegraphics[clip,width=8.6cm]{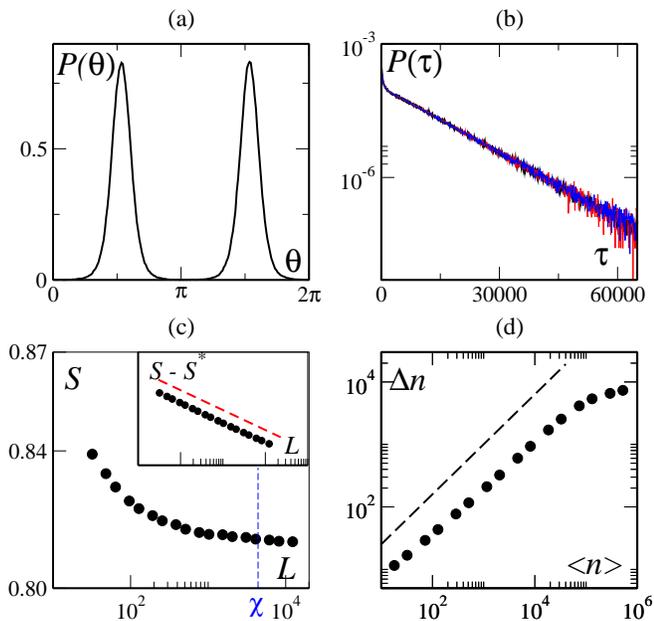} 
\caption{(color online) Phase I (homogeneous nematic order, $\eta=0.095$). 
(a) Polar orientation probability distribution in a system of size $L=2048$.   
(b) Distribution of particles transition times $\tau$ between the two peaks of (a) for
three different system sizes $L=512$, 1024, and 2048 
(black, red, and blue lines respectively).
(c) Nematic order parameter $S$ vs
system size $L$ in square domains. The vertical red dashed line marks the persistence length  
$\chi \approx 4400$ (see text).  
Inset: $S-S^*=0.813063$ vs $L$ (red dashed line: $L^{-2/3}$ decay). 
(d) Number fluctuations $\Delta n$ as a function of average particle number 
$\langle n \rangle$ (see text) in a system of size $L=4096$  
(dashed line: algebraic growth with exponent $0.8$).} 
\label{fig:4} 
\end{center} 
\end{figure}

Phase I, present at the lowest $\eta$ values, is ordered and  
spatially homogeneous (Fig.~\ref{fig:2}a). Its nematic order, 
which arises quickly from any initial condition, is due to the existence, 
at any time, of two subpopulations of particles
that migrate in opposite directions (Fig.~\ref{fig:4}a). 
Statistically of equal size, they constantly exchange particles, 
those which ``turn around''. These events   
occur at exponentially-distributed times $\tau$ (Fig.~\ref{fig:4}b). 
Increasing system size, the nematic order parameter $S$ is almost constant, 
decaying {\it slower than a power law} (Fig.~\ref{fig:4}c). 
A good fit of this decay is given 
by an algebraic approach to a constant asymptotic value $S^*$. Thus, 
our data seem to indicate the existence of true long-range nematic order.
(Quasi-long-range order, expected classically for two-dimensional nematic phases,
is characterized by an algebraic decay of $S$.)
A discussion of this striking fact is given below. 
Finally, as expected on general grounds for homogeneous ordered phases of  
active particles \cite{Ramaswamy2003}, phase I exhibits so-called 
giant number fluctuations: the fluctuations 
$\Delta n^2=\langle (n-\langle n\rangle)^2\rangle$ of the 
average number of particles $\langle n \rangle = \rho \ell^2$ contained 
in a square of linear size $\ell$ follow the power law 
$\Delta n \sim \langle n \rangle ^\alpha$ with $\alpha>\frac{1}{2}$ (Fig.~\ref{fig:4}d).
Our estimate of $\alpha$ is compatible to that measured for {\it polarly} 
ordered phases $\alpha=0.8$ \cite{Chate2008}.

\begin{figure}[t!] 
\begin{center} 
\includegraphics[clip,width=8.6cm]{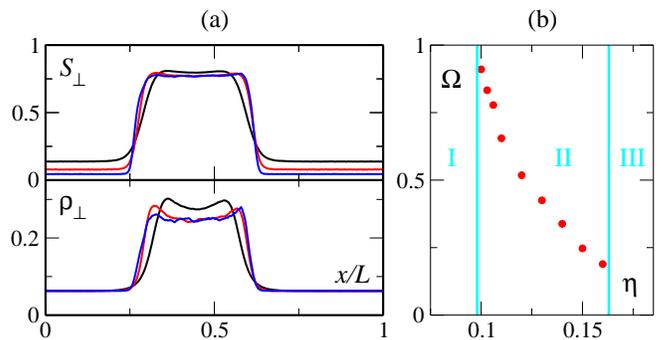} 
\caption{(color online) Phase II (stable bands)
(a) Rescaled transverse profiles in square domains of linear size 
$L=512$ (black), $1024$ (red), and $2048$ (blue) at $\eta=0.14$.
(Data averaged over the longitudinal direction and time, translated to be centered
at the same location.)
Bottom: density profiles. Top: nematic order parameter profiles. 
(b) Surface fraction $\Omega$ as a function of noise amplitude $\eta$ (defined here as
the width at mid-height of the rescaled $S$ profile).}
\label{fig:5} 
\end{center} 
\end{figure}

Phase II differs from phase I by the presence, in the steady-state, 
of a low-density disordered region. In large-enough systems ($10^4$-$10^5$ particles for
the parameters used here), 
a narrow, low density channel emerges (Fig.~\ref{fig:2}b) 
when increasing $\eta$ from phase I. It becomes wider at larger $\eta$ values,  
so that one can then speak of a high-density ordered band, 
typically oriented along one of the main axes of the box, amidst a  
disordered sea (Fig.~\ref{fig:2}c).  
Particles travel along the high-density band, turning around or leaving the band
from time to time.
Within the band, nematic order with properties similar to those of phase I 
is found (slow decay of $S$ with system size, giant number fluctuations).
The (rescaled) band possesses 
a well-defined profile with sharper and sharper edges as $L$ increases 
(Fig.~\ref{fig:5}a). The fraction area $\Omega$ occupied by the band is 
thus asymptotically independent of system size, and it decreases 
continuously as the noise strength $\eta$ increases (Fig.~\ref{fig:5}b). 
 
\begin{figure}[t] 
\begin{center} 
\includegraphics[clip,width=8.6cm]{Fig6.eps} 
\includegraphics[clip,width=8.6cm]{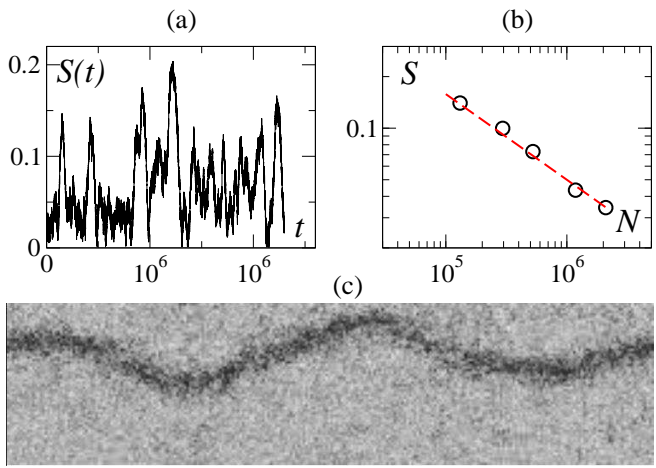} 
\caption{Phase III (unstable bands, $\eta=0168$). 
(a) Typical nematic order parameter time series for a system of linear  size $L=2048$.  
(b) $S$ vs $N$ in square domains of increasing sizes.  
(The dashed line marks a $1/\sqrt{N}$ decay.)
(c) Snapshot of coarse-grained density field during the growth of the instability of 
an initially straight band in a $2048\times 512$ domain.} 
\label{fig:6} 
\end{center} 
\end{figure}

In phase III, spontaneous segregation into bands still occurs  
(for large-enough domains), however these thinner bands are unstable 
and constantly bend, break, reform, and merge, in an  
unending spectacular display of space-time chaos (Fig.~\ref{fig:2}d) \cite{EPAPS}.
Correspondingly, $S(t)$ fluctuates strongly (Fig.~\ref{fig:3}) and on very large time 
scales (Figs.~\ref{fig:6}a). Nevertheless, these fluctuations behave normally 
({\it i.e.} decrease like $1/\sqrt{N}$, Fig.~\ref{fig:6}b). Thus, the   
space-time chaos self-averages, making phase III a bona fide disordered 
phase, albeit one with huge correlation lengths and times. 
 
Phase IV, observed for the highest noise strengths, exhibits local and global disorder on small 
length- and time-scales, and is spatially homogeneous (Fig.~\ref{fig:2}e). 
 
We now discuss briefly the nature of the three transitions that separate   
the four observed phases (details will appear elsewhere \cite{TBP}) .  
The I--II transition, located at $\eta_{\rm I-II}\simeq 0.098(2)$,  
is characterized by the emergence of a narrow  
low-density disordered channel. Within phase II, the 
emergence of these structures from disordered initial conditions
is reminiscent of a nucleation 
process. Even though the emerging channels might occupy an arbitrarily small 
proportion of space near the transition ($\Omega\sim 1$  
for $\eta\gtrsim\eta_{\rm I-II}$), they seem to possess  
a minimum absolute width. 
These facts suggest a discontinuous I--II transition. 
The transition between phase II and III, located near  
$\eta_{\rm II-III}\simeq 0.163(1)$, constitutes the order-disorder transition of the model. 
As mentioned above, it resembles a long-but-finite wavelength instability of the band 
(see, e.g., Fig.~\ref{fig:6}c)  
and does not appear as a fluctuation-driven phase transition. 
The disorder-disorder transition between phases III and IV occurs near
$\eta_{\rm III-IV}\simeq 0.169(1)$, where the instantaneous order parameter $S(t)$ 
exhibits a bistable behavior between a low value, fast fluctuating state typical of phase IV  
and a larger amplitude, slowly fluctuating one characteristic of phase III.  
This bistability, leading to a bimodal order parameter distribution, 
suggests a discontinuous phase transition.

At this point, the most crucial question is perhaps
that of the stability of the nematic order observed in phases I and II.  
Indeed, much of what we described above for large but finite systems relies on 
our conclusion of possible truly long-range (asymptotic) order (Fig.~\ref{fig:4}c). 
On the one hand, one could argue that the exponential distributions 
of flight times between the two opposite polar orientations (Fig. \ref{fig:4}b) define 
a finite persistence time $\tau$ and a corresponding 
finite persistence lengthscale $\chi \approx v_{\rm 0} \tau$ 
(indicated by the blue dashed line in Fig.~\ref{fig:4}c). 
Therefore, at scales much larger than $\chi$, the polar nature of 
our particles could become irrelevant, and the system would then behave like a fully nematic one,
with only quasi-long-range order.
As of now, we have been able to probe systems sizes up to three or four times the persistence
length $\chi$. So far, as shown in Fig.~\ref{fig:4}c, 
these systems comprising up to twenty million particles show no sign of breakdown of order.
On the other hand,  $\chi$ is a single-particle quantity. Even though 
it is finite and system size independent, particles travel in rather  
dense {\it polar} packets which have flights longer than $\chi$.
Indeed, the giant density fluctuations 
reported (Fig.~\ref{fig:4}d) indicate that denser, more ordered, and hence  
probably longer-lived packets occur in larger systems. 
Unfortunately, packets' flight times are hard to define and measure \cite{TBP}.
But should this ``polar packet lifetime'' diverge with system size, then one would have 
a mechanism opening the door for the emergence of true long-range nematic order.
To summarize this discussion, nematic order could break down for sizes much larger than
$\chi$, but our data (Figs.~\ref{fig:4}c,d) and the argument above suggest 
the picture of two opposite polar components each with true long-range order 
(as in fully-polar models \cite{NOTEGNF}) summing up to true nematic order.

Further work is thus needed, but most of our results are rather robust. 
For instance, the introduction of some soft-core short-range repulsion between particles 
does {\it not} modify our main findings \cite{TBP}. Thus, these are not due 
to the pointwise nature of the particles, and should also be observed in 
previous, more detailed models of self-propelled rods if sufficiently-large populations 
are considered.
We note also that our results, and in particular the instability and space-time chaotic 
motion of the spontaneously segregated bands (phases II and III) \cite{EPAPS}, 
are reminiscent of 
the streaming and swirling regime which characterizes the aggregation of 
myxobacteria \cite{myxo,myxo2} and thus our model could prove relevant in this context.
 
At a more general level, our findings reveal unexpected emergent behavior among even 
the simplest situations giving rise to collective motion. 
Our model of self-propelled polar objects
aligning nematically stands out as a member of a universality class
distinct from both that of the Vicsek model \cite{Vicsek1995,Toner,Chate2008} 
and of active nematics \cite{Chate2006}.
Thus, in this out-of-equilibrium context, the symmetries of
the moving particles and of their interactions must be considered separately
and are both relevant ingredients.

We thank J. Toner and S. Ramaswamy for fruitful discussions. 
This work was partially funded by the French ANR projects ``Morphoscale'' and ``Panurge'', and the German DFG grants DE842/2, SFB 555, and GRK 1558.


\begin{thebibliography}{99}
\bibitem{NOTE1} The reverse case of active nematic particles with a 
ferromagnetic interaction does not make sense.
\bibitem{NOTEGNF} This is also suggested by our observation that the density fluctuations
are governed by the same exponent as in the polar case (Fig.~\ref{fig:4}d).
\end{thebibliography}
\end{document}